\documentclass[aps,prl,twocolumn,superscriptaddress,showpacs,floatfix]{revtex4-1}
\usepackage{graphicx}
\usepackage{amsmath}
\usepackage{amssymb}
\usepackage{url}
\usepackage{float}
\usepackage{color}
\usepackage{natbib}

\begin{document}

\title{Quasiparticle renormalization in ABC graphene trilayers}

\author{Xu Dou}
\affiliation{Department of Physics and Astronomy, University of Oklahoma, OK 73069,
USA}
\author{Akbar Jaefari}
\affiliation{Department of Physics and Astronomy, University of Oklahoma, OK 73069,
USA}
\author{Yafis Barlas}
\affiliation{Department of Physics and Astronomy, University of California at
Riverside, CA 92521, USA}
\author{Bruno Uchoa}
\affiliation{Department of Physics and Astronomy, University of Oklahoma, OK 73069,
USA}

\date{\today}
\begin{abstract}
We investigate the effect of electron-electron interactions in ABC
stacked graphene trilayers. In the gapless regime, we show that the
self-energy corrections lead to the renormalization of the of dynamical
exponent $z=3+\alpha_{1}/N$, with $\alpha_{1}\approx0.52$ and $N$
is the number of fermionic species. Although the quasiparticle residue
is suppressed near the neutrality point, the lifetime has a sublinear
scaling with the energy and the quasiparticles are well defined even
at zero energy. We calculate the renormalization of a variety of physical
observables, which can be directly measured in experiments. 
\end{abstract}


\maketitle
\emph{Introduction.} In graphene single layers, the honeycomb arrangement
of the carbon atoms leads to a linear electronic dispersion and to
quasiparticles that behave as massless Dirac fermions, akin to massless
neutrinos in quantum electrodynamics (QED) \cite{grapheneRMP,Kotov}.
In graphene multilayers, the electronic spectrum varies depending
on the stacking sequence. In the single particle picture, rombohedral
ABC-stacked trilayer graphene reveals a gapless band structure of
chiral quasiparticles with Berry phase 3$\pi$ and \emph{cubic} low
energy excitation spectrum \cite{Guinea,Zhang}. Because of the scaling
of the kinetic energy, Coulomb interactions are relevant operators
in the renormalization group (RG) sense, and can strongly renormalize
different physical quantities. Different spontaneous broken symmetry
ground states have been already proposed for trilayer graphene \cite{Cvetovic,Gorbar,Olsen}.
Very recently, transport experiments revealed a robust many-body gap
of $\sim$40 meV at temperatures below $T_{c}\sim34$K \cite{Lee}. 

In this letter we study the effect of Coulomb interactions and polarization
effects on the behavior of the quasiparticles at small but finite
temperature, when the many-body gap is zero. We investigate the analytical
structure of the polarization bubble and the leading self-energy corrections
due to \emph{dynamically} screened Coulomb interactions. In the gapless
regime, we show that the dynamical critical exponent is renormalized
to 
\[
z=3+\alpha_{1}/N+O(N^{-2}),
\]
where $\alpha_{1}\approx0.52$ and $N=4$ is the number of fermionic
flavors. Although the quasiparticle residue is suppressed by interactions,
the scattering rate has a sublinear scaling with energy and the quasiparticles
remain well defined. We predict the renormalization of several physical
observables in the metallic phase, such as the electronic compressibility,
the specific heat, the density of states (DOS) and the spectral function,
which can be measured with angle resolved photoemission (ARPES) experiments.

\emph{Low energy Hamiltonian.} We start with a simplified two-band
model where the high energy bands are separated in energy by interlayer
hopping processes, which set the ultraviolet cut-off for the excitations
in the low-energy bands, $t_{\perp}\sim0.4$eV. We will assume a temperature
regime above the ordering temperature $T\gtrsim T_{c}\sim4$ meV, where
the band structure is gapless. The infrared cut-off of the model is
the trigonal warping energy $\sim10$ meV, below which the bands disperse
quadratically \cite{Zhang}. 

The low energy physics of the non-interacting ABC-trilayer in the
gapless regime is described by the $2\times2$ Hamiltonian $\mathcal{H}_{0}=\sum_{\mathbf{p}}\Psi_{\mathbf{p}}^{\dagger}\hat{\mathcal{H}}_{0}(\mathbf{p})\Psi_{\mathbf{p}}$,
where $\Psi_{\mathbf{k}}=(a_{,\mathbf{k}},\bar{b}_{,\mathbf{k}})$
is a two component spinor defined in terms of one annihilation operator
in sublattice $A$ of the top layer ($a_{\mathbf{p}})$ and another
in sublattice $B$ for the bottom layer (\textbf{$\bar{b}_{\mathbf{p}}$}).
The total degeneracy is $N=4$, including spin and valley degrees
of freedom. The Hamiltonian density operator is \cite{Guinea,Zhang}
\begin{equation}
\hat{\mathcal{H}}_{0}=\frac{(\hbar v)^{3}}{t_{\perp}^{2}}\left(\begin{array}{cc}
0 & (\pi)^{3}\\
(\pi^{\dagger})^{3} & 0\end{array}\right),\label{eq:Ho}
\end{equation}
where $\hbar v\approx6$ eV$\mbox{\AA}$ is the Fermi velocity, and
$\pi=p_{x}-ip_{y}$ is defined by the $x$ and $y$ components of
the in-plane momentum of the quasiparticles measured away from the
neutrality point. In a more compact notation, $\hat{\mathcal{H}}_{0}(\mathbf{k})=\gamma|\mathbf{k}|^{3}\hat{h}_{0}(\mathbf{k})$
with \begin{equation}
\hat{h}_{0}(\mathbf{k})=\mathrm{cos}(3\theta_{\mathbf{k}})\sigma^{1}+\mathrm{sin}(3\theta_{\mathbf{k}})\sigma^{2},\label{eq:h}
\end{equation}
where $\sigma^{i}$ ($i=1,2$) are $x,\, y$ Pauli spin matrices,
and $\mathrm{tan}\theta_{\mathbf{k}}=k_{y}/k_{x}$. The constant $\gamma\equiv(\hbar v)^{3}/t_{\perp}^{2}$,
is proportional to the velocity of the quasiparticles $\mathbf{v}_{0}=\partial_{\mathbf{k}}E_{\mathbf{k}}$,
which have the energy spectrum $\pm E_{\mathbf{k}}=\pm\gamma|\mathbf{k}|^{3}$. 

In ABC trilayers, Coulomb interactions are relevant in the RG flow
at the tree level, and hence standard perturbation theory is not possible.
We organize the expansion of the self-energy corrections in powers
of the dynamically screened Coulomb interaction, which can be rigorously
justified in the large $N$ limit. At long wavelengths, $k\ll1/d$,
where $d\sim2.4\mbox{\AA}$ is the interlayer distance, the bare Coulomb
interaction is 
\begin{equation}
\mathcal{H}_{I}=\frac{1}{2}\sum_{\mathbf{q}}\, V(q)\hat{n}(\mathbf{q})\hat{n}(-\mathbf{q}),\label{eq:Hi}
\end{equation}
with $\hat{n}(\mathbf{q})$ a density operator and $V(q)\approx2\pi e^{2}/q$,
as in a 2D system. In the long wavelength regime where this approximation
is valid, the DOS scales as $\rho(\mathbf{q})=(6\pi\gamma)^{-1}/q$
and the screened Coulomb interaction is $\tilde{V}(q,\omega)=V(q)/\left[1-V(q)\Pi(\mathbf{q},\omega)\right],$
where $\Pi(\mathbf{q},\omega)$ is the dynamical polarization function.
In trilayers, the large $N$ approximation becomes asymptotically
exact at small momentum, where the DOS diverges and screening becomes
strong. 

\begin{figure}[t]
\includegraphics[scale=0.36]{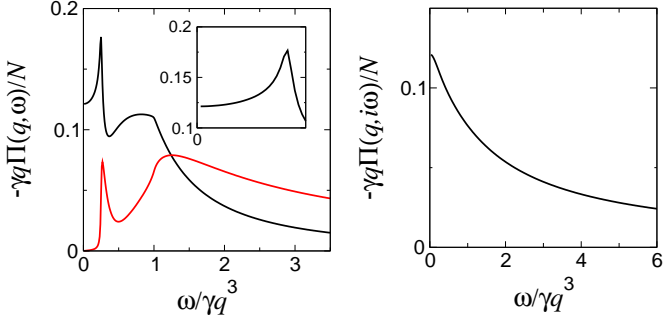} \label{fig:polarizationfunctions}
\vspace{-0.2cm}
%\par\end{centering}
\caption{{\small Left: Polarization bubble in one loop calculated numerically
from Eq. (\ref{eq:f2-1}). The real part (black curve) has a logarithmic
singularity at the edge of the particle-hole continuum, at $\omega=\gamma q^{3}/4,$
shown in detail in the inset. Red curve: imaginary part. Right panel:
Polarization in imaginary frequencies, which is a purely real function.
For $\omega/\gamma q^{3}\gg1$, $\Pi^{(0)}(q,i\omega)\to-3Nq^{2}/(16\omega)$
(see text).}}
\end{figure}

\emph{Polarization bubble.} In order to address the screening effects,
we consider the bare polarization function, which is defined as $\Pi^{(0)}(\mathbf{q},\omega)=\frac{1}{\beta}\mbox{tr}\sum_{i\nu}\sum_{\mathbf{p}}\hat{G}_{0}(\mathbf{p},i\nu)\hat{G}_{0}(\mathbf{p}+\mathbf{q},i\omega+i\nu),$
where
\begin{align}
\hat{G}_{0}(\mathbf{q},i\omega) & =\frac{1}{2}\sum_{s=\pm}\frac{1+s\hat{h}_{0}(\mathbf{q})}{i\omega-s\gamma q^{3}}
\end{align}
is the fermionic Greens function, described by a 2$\times2$ matrix.
After performing the sums over the Matsubara frequencies, the polarization
function is given by
\begin{align}
\Pi^{(0)}(\mathbf{q},\omega)=-\frac{N}{2}\int\frac{d^{2}p}{(2\pi)^{2}}\sum_{s=\pm}\frac{1-\cos(3\theta_{{\bf p}{\bf q}})}{E_{{\bf p}+{\bf q}}+E_{{\bf p}}-s\omega}\label{eqn:polarization1}
\end{align}
 where $\theta_{{\bf p}{\bf q}}=\theta_{{\bf p}+{\bf q}}-\theta_{{\bf p}}$
is the angle between vectors ${\bf p}+{\bf q}$ and ${\bf p}$. By
sending the ultraviolet cut-off to infinity, a simple dimensional
analysis reveals the functional form of the polarization function
to be $\gamma q\Pi^{(0)}({\bf q},i\omega)=-Nf(i\omega/(\gamma q^{3})).$
After some algebra, the scaling function $f(z)$ can be written in
the form 
\begin{align}
f(iz)= & \,\frac{1}{2}\int_{0}^{2\pi}\!\!\mathrm{d}\theta\!\int_{0}^{\infty}\!\!\frac{\mbox{d}x\, x}{(2\pi)^{2}}\sum_{s=\pm}\frac{s}{iz+s\left[x^{3}+h^{3}(x,\theta)\right]}\quad\nonumber \\
 & \,\times\left[1-4\left(\frac{1+x\mathrm{cos}\theta}{h(x,\theta)}\right)^{3}+3\left(\frac{1+x\mathrm{cos}\theta}{h(x,\theta)}\right)\right],\label{eq:f2}
\end{align}
where $z=\omega/(\gamma q^{3})$ and $h(x,\theta)\equiv\sqrt{1+x^{2}+2x\mathrm{cos}\theta}$.
$f(z)$ is a well-defined function in imaginary frequency but has
branch cuts related to the edge of the particle-hole continuum on
the real axis. Due to the cubic dispersion, it is difficult to come
up with a closed form solution for the polarization function. However
the analytical structure of $f(z)$ near the particle-hole threshold
$z=1/4$ can be extracted in the collinear scattering approximation,
which dominates the processes near that region \cite{Gangadharaiah}.
We consider the singular contribution of the integrand around the
momenta $\mathbf{p+q}\approx-\mathbf{p}$. Within this window it is
safe to assume $1-\cos(3\theta_{{\bf p}{\bf q}})\approx2$. After
expanding $\cos\theta$ around $\theta=\pi$ to the second order,
we arrive at the following integral representation for $f(z)$,
\begin{align}
f(z) & \cong\int\frac{xdx}{(2\pi)^{2}}\int\frac{d\theta}{x^{3}+(1-x)^{3}+\frac{3}{2}x(1-x)\theta^{2}-z}.\label{eq:f2-1}
\end{align}
Considering the rapid fall of the integrand with respect to $\theta$
around $\pi$, one can conveniently extend the upper limit of the
angular integral to infinity, $\theta\in[0,\infty[$. After performing
the integrals, we arrive at the most dominant part of $f(z)$ near
$z\sim1/4$,
\begin{equation}
f(z)=-\frac{1}{6\sqrt{2}\pi}\ln\left(1-4|z|\right)+\text{regular terms},\label{eq:f3}
\end{equation}
which describes a logarithmic divergence near the edge of the particle
hole continuum. Exploring the two asymptotic regimes, in the $z\to0$
regime, $f(0)=c_{0}\approx0.12$ is a constant \cite{Min,Gelderen}
and in the $z\gg1$ limit, $f(z)\to-ic_{\infty}/z$ is purely imaginary,
with $c_{\infty}=3/16$. 

In Fig. 1, we show the behavior of the real and imaginary parts of
$f(z)$ calculated numerically from Eq. (\ref{eq:f2-1}). The scaling
function has only one singularity near $z\sim1/4$. For $z<1/4$,
$f(z)$ is purely real and diverges logarithmically at $z=1/4$, in
agreement with the analytical expression (\ref{eq:f3}), as shown
in the inset of Fig. 1. For $z>1/4$, $f(z)$ has also an imaginary
part, which decays with $1/z$. The right panel of Fig. 1 shows $f(iz)$
in imaginary frequency, which is a real and well behaved monotonic
function.

In the optical regime, for $z\gg1$, where $\Pi^{(0)}(q,\omega)\to iNc_{\infty}q^{2}/\omega$,
the optical conductivity can be calculated directly from the charge
polarization, 
\begin{equation}
\sigma(\omega)=\frac{e^{2}}{\hbar}\lim_{q\to0}\frac{i\omega}{q^{2}}\frac{\Pi^{(0)}(\mathbf{q},\omega)}{1-V(q)\Pi^{(0)}(\mathbf{q},\omega)}=\frac{3}{4}\frac{e^{2}}{\hbar},\label{eq:sigma}
\end{equation}
which is proportional to the Berry phase $3\pi$. In the general case,
$\sigma(\omega)=\nu e^{2}/(2h)$, with $\nu=\pi$ for graphene single
layer and $\nu=2\pi$ for bilayers. 

\emph{Self-energy.} The leading self energy correction due to the
screened Coulomb interaction is diagrammatically shown in Fig. 2.
In imaginary time, the self-energy is given by 
\begin{equation}
\hat{\Sigma}^{(1)}(\mathbf{q},i\omega)=-\frac{1}{\beta}\sum_{\nu}\int\!\!\frac{\mbox{d}^{2}p}{(2\pi)^{2}}\tilde{V}(\mathbf{p},i\nu)\hat{G}^{(0)}(\mathbf{q}-\mathbf{p},i\omega-i\nu).\label{eq:Sigma}
\end{equation}
Through power counting, the leading divergences appear at long wavelengths,
where the large $N$ limit is a good approximation. At large $N$,
the dynamically screened potential is approximated by $\tilde{V}({\bf q},i\omega)\approx\gamma q/[Nf(i\omega/\gamma q^{3})]+O(N^{-2})$
\cite{Son,Foster,Nmass}. Since $f(iz)$ is a well behaved function,
with no singularities or branch cuts, the self energy in one loop
can be calculated directly in the zero temperature limit. The leading
contribution is logarithmically divergent, 
\begin{equation}
\Sigma^{(1)}(\mathbf{q},i\omega)=\frac{1}{2\pi^{2}N}\left[\alpha_{d}i\omega+\alpha_{o}\gamma q^{3}\hat{h}(\mathbf{q})\right]\mathrm{ln}\left(\frac{\Lambda}{q}\right),\label{eq:Self}
\end{equation}
where $t_{\perp}=\gamma\Lambda^{3}$ defines the ultraviolet cut-off
in momentum, namely $\Lambda=t_{\perp}/(\hbar v)$. The coefficients
\begin{equation}
\alpha_{o}=\int_{0}^{\infty}dz\frac{1}{f(iz)}\frac{z^{2}(10-16z^{2}+z^{4})}{(1+z^{2})^{4}},\label{eq:ctilde}
\end{equation}
and 
\begin{equation}
\alpha_{d}=\int_{0}^{\infty}dz\frac{1}{f(iz)}\frac{1-z^{2}}{(1+z^{2})^{2}},\label{eq:alpha_d}
\end{equation}
can be found though numerical integration using the exact $f(iz)$
from Eq. (\ref{eq:f2-1}). Although $\alpha_{o}$ and $\alpha_{d}$
both diverge logarithmically with the upper limit of integration at
large $z$, we will postpone their regularization for the moment,
since these divergences cancel exactly in the renormalization of $\gamma$
and hence have no consequence to the renormalization of the spectrum. 

The self-energy can be separated in two terms, $\hat{\Sigma}(\mathbf{q},i\omega)=i\omega\Sigma_{d}\sigma_{0}+\Sigma_{0}q^{3}\hat{h}_{0}(\mathbf{q}),$
where $\Sigma_{d}$ is the diagonal term, and $\Sigma_{o}$ describes
the off-diagonal matrix elements. The diagonal part of the self-energy
has frequency dependence and defines the quasiparticle residue renormalization,
\begin{equation}
Z_{\psi}^{-1}=1-\partial\hat{\Sigma}/\partial(i\omega)=1-\Sigma_{d}.\label{Z}
\end{equation}
The renormalized Green's function is $\hat{G}(\mathbf{q},i\omega)=Z_{\psi}[i\omega-\gamma\hat{h}_{0}(\mathbf{q})Z_{\psi}(1+\Sigma_{o})]^{-1}$.
In one loop, the renormalized energy spectrum is 
\begin{equation}
\frac{\gamma(q)}{\gamma}=\frac{1+\Sigma_{o}}{1-\Sigma_{d}}\approx1-\frac{\alpha_{1}}{N}\ln\!\left(\frac{\Lambda}{q}\right)+O(1/N^{2}),\label{eq:gamma ren}
\end{equation}
where 
\begin{equation}
\alpha_{1}=\frac{\alpha_{0}+\alpha_{d}}{2\pi^{2}}=\int_{0}^{\infty}\frac{\mbox{d}z}{2\pi^{2}}\frac{1}{f(iz)}\frac{17z^{4}-11z^{2}-1}{(1+z^{2})^{4}}\approx0.52\label{alpha1}
\end{equation}
is a finite well defined quantity. 

\begin{figure}[t]
\vspace{-0.5cm}\includegraphics[scale=0.35]{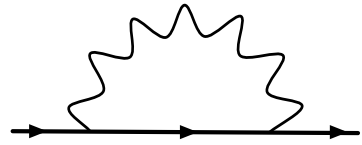}\vspace{-0.2cm}\vspace{-0.3cm}
 \label{fig:sunshine} %\par\end{centering}
\caption{{\small One-loop correction to the self-energy with the dressed Coulomb
interaction.}}
\end{figure}

The logarithmic renormalization of the quasiparticle velocity in one
loop dictates the RG equation of $\gamma$, 
\begin{equation}
\beta_{\gamma}\equiv\frac{\mathrm{d}\gamma}{\mathrm{d}l}=-\gamma\frac{\alpha_{1}}{N},
\end{equation}
where $l=\ln(\Lambda/\Lambda^{\prime})$, with $\Lambda^{\prime}<\Lambda$
the renormalized cut-off, whose solution is 
\begin{equation}
\gamma(q)=\gamma\times[(\hbar v/t_{\perp})q]{}^{\alpha_{1}/N}.\label{eq:gamma3}
\end{equation}
The energy spectrum acquires an anomalous dimension $\eta=\alpha_{1}/N$,
which leads to the renormalization of the dynamical exponent, $\omega\propto q^{z}$,
with $z=3+\alpha_{1}/N+O(N^{-2})$. This result can be related with
the graphene bilayer case, where $\eta=0.078/N$ \cite{Lemonik} and
with the large $N$ limit of the single layer case, where $\eta=-4/(\pi^{2}N)$
\cite{Son,Foster}.

This analysis can be explicitly verified by checking the two loop
correction in the self energy. The RG equation describes a resummation
of leading logs to all orders in $1/N$. The $N^{-2}\log^{2}$ terms
cancel exactly in the vertex correction diagram at two loop, and hence
vertex corrections do not renormalize in the RG flow \cite{Lemonik}.
The leading logarithmic terms appear in the remaining diagrams of
the same order, and lead to a second order correction to Eq. (\ref{eq:gamma ren}),
$\gamma^{(2)}(q)/q=\frac{1}{2}\alpha_{1}^{2}/N^{2}\ln^{2}(\Lambda/q)$,
in agreement with the result of the RG equation up to $1/N^{2}$ order. 

\emph{Quasiparticle residue}. To calculate the quasiparticle residue
renormalization $Z_{\psi}$ through Eq. (\ref{Z}), one needs to regularize
integral (\ref{eq:alpha_d}). That can be done introducing an upper
cut-off $z_{c}$ which accounts for the condition where the large
$N$ limit breaks down, namely $-V(p)\Pi^{(0)}(p,i\nu)=2\pi Ne^{2}\Lambda^{2}/(\hbar vp^{2})f(iz_{c})\sim1.$
At large $z$, where $f(iz)\to3/(16z)$, the leading contribution
is $\alpha_{d}\sim-16\ln(\Lambda/p)$. Replacing $\ln(\Lambda/q)\to\int_{q}^{\Lambda}\mbox{d}p/p$
in Eq. (\ref{eq:Self}) and carrying out the momentum integration,
the quasiparticle residue $Z_{\psi}$ is given by 
\begin{equation}
Z_{\psi}^{-1}\to1+\frac{4}{\pi^{2}N}\ln^{2}(\Lambda/q)+O(1/N^{2}),\label{Z-1}
\end{equation}
in one loop, and is suppressed logarithmically in the infrared. 

In the RG spirit, we now reestablish the bare value of the quasiparticle
residue $Z_{0}$ in the bare Green's function $\hat{G}_{0}\propto Z_{0}$
\cite{Nandkishore}, and set $Z_{0}\to1$ at the end. Since $\delta\hat{G}=\hat{G}_{0}\hat{\Sigma}\hat{G}_{0}\propto\delta Z_{\psi}$
in lowest order in the Dyson equation, then $\delta Z_{\psi}=Z_{0}^{2}\hat{\Sigma}_{d}\propto Z_{0}$
in large $N$. Eq. (\ref{Z-1}) then becomes $\delta Z_{\psi}=-4Z_{0}/(\pi^{2}N)\delta\ln^{2}(\Lambda/q)$,
which corresponds to the RG equation 
\begin{equation}
\beta_{\psi}=\frac{\mbox{d}Z_{\psi}}{\mbox{d}l}=-\frac{8}{\pi^{2}N}lZ_{\psi},\label{eq:betapsi}
\end{equation}
with $l=\ln(\Lambda/\Lambda^{\prime})$, whose solution is 
\begin{equation}
Z_{\psi}(q)=\mbox{exp}\left[-4/(\pi^{2}N)\ln^{2}(\Lambda/q)\right],\label{RG sol}
\end{equation}
in agreement with Eq. (\ref{Z-1}) up to $1/N$ order. 

\begin{figure}[t]
\includegraphics[scale=0.35]{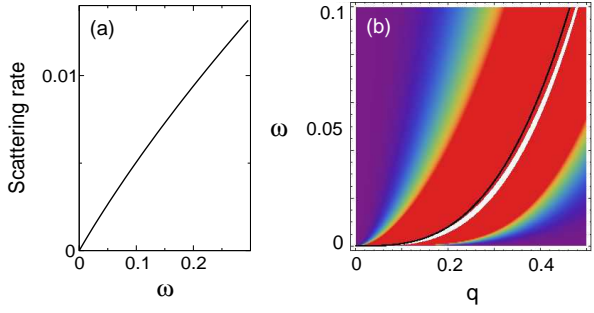}\hspace{4cm}\vspace{-0.1cm}
 \label{fig:sunshine-1} %\par\end{centering}
\caption{{\small a) On-shell scattering rate $\tau(\omega)$ vs energy in units
of $t_{\perp}\sim0.4$eV. b) Density plot of the spectral function
as a function of energy $(\omega/t_{\perp})$ and momentum ($q/\Lambda$).
Solid black line: bare energy spectrum. White line: renormalized one.
Light regions indicate higher intensity. }}
\end{figure}

\emph{Quasiparticle lifetime.} In real frequency, the polarization
function has a logarithmic branch cut. To calculate the quasiparticle
scattering rate $\tau=Z_{\psi}\mbox{Im}\hat{\Sigma}$, we use the
method in ref. {[}\onlinecite{QuinnFerrell}{]} to separate the
self-energy into the line part and the residue part, $\hat{\Sigma}=\hat{\Sigma}_{\text{line}}+\hat{\Sigma}_{\text{res}}.$
The line part is obtained by performing Wick rotation $i\omega\to\omega+i0_{+}$
in the self-energy (\ref{eq:Sigma}), and is purely real. The residue
part follows from the residue calculated around the pole of the Green's
function, 
\begin{align}
 & \Sigma_{\text{res}}^{(1)}({\bf q},\omega)=-\frac{1}{2}\sum_{s=\pm}\int\frac{\mathrm{d}^{2}p}{(2\pi)^{2}}\tilde{V}(|{\bf q}|,\omega)[1+s\hat{h}(\mathbf{q}-\mathbf{p})]\nonumber \\
 & \quad\qquad\times[\theta(\omega-s\gamma|{\bf q}-{\bf p}|^{3})-\theta(-s\gamma|{\bf q}-{\bf p}|^{3})],
\end{align}
with $\theta$ a step function. The scattering rate is given by $\tau(\mathbf{q},\omega)=Z_{\psi}\mbox{Im}\Sigma_{\mathrm{res}}(\mathbf{q},\omega)$.
In the on-shell region, near $\omega\sim\gamma q^{3}$, $\tau(\omega)=\omega Z_{\psi}g(\omega/t_{\perp}),$
where
\begin{equation}
g(y)=\frac{1}{2N}\,\mbox{Im}\!\int_{|\mathbf{x}|<1}\frac{\mbox{d}^{2}x}{(2\pi)^{2}}\frac{|\hat{q}-\mathbf{x}|}{\bar{\alpha}y^{2/3}|\hat{q}-\mathbf{x}|^{2}+f\!\left(\frac{1-x^{3}}{|\hat{q}-\mathbf{x}^{3}|}\right)}\label{eq:g}
\end{equation}
is a scaling function in one loop, with $y=\omega/t_{\perp}$, $\bar{\alpha}=\hbar v/(2\pi Ne^{2})$
is a dimensionless constant and $\hat{q}=\mathbf{q}/q$. The function
$g(y)$ has a very slow variation, as shown in Fig. 3a, and as a consequence
$\tau(\omega)\sim\omega Z_{\psi}[(\omega/\gamma)^{1/3}]$ has a sublinear
scaling with energy within logarithmic accuracy. In the large $N$
limit ($\bar{\alpha}\to0$), which is valid at low energy, $g(y)\approx0.043$
is a constant. Since the ratio $\tau(\omega)/\omega\ll1$, the quasiparticles
are well defined even in the $\omega\to0$ limit.

The spectral function is given by $A(\mathbf{q},\omega)=-2\mbox{tr}\,\mbox{Im}G^{R}(\mathbf{q},\omega)$,
where 
\begin{equation}
\hat{G}^{R}(\mathbf{q},\omega)=\frac{1}{2}\sum_{s=\pm}\frac{Z_{\psi}(q)[1+s\hat{h}_{0}(\mathbf{q})]}{\omega-s\gamma(q)q^{3}-i\tau(\omega)+i0^{+}}\label{G}
\end{equation}
 is the retarded part of the renormalized Green's function. The spectral
function is shown in Fig. 3b. The solid black line describes the bare
energy spectrum, while the light region describes the renormalized
one, which corresponds to the pole of the renormalized Green's function.
There is a clear deviation of the two curves, which could be observed
with ARPES experiments.

\emph{Other physical observables. }The renormalization of the quasiparticle
velocity encoded in the RG flow of $\gamma$ leads to the renormalization
of many physical observables. For instance, the specific heat for
non-interacting particles with cubic dispersion in 2D scales with
$C\sim(T/\gamma)^{2/3}$, where $T$ is the temperature. From Eq.
(\ref{eq:gamma3}), the scaling of $\gamma$ with energy is $\gamma\sim\omega^{\alpha_{1}/(3N)}$.
At $\omega\sim T$, the temperature scaling of the specific heat is
renormalized to 
\begin{equation}
C\sim T^{2(1-\alpha_{1}/(3N)]/3}\approx T^{2/3-0.1/N},\label{C}
\end{equation}
neglecting slower logarithmic corrections due to the scaling of $Z_{\psi}$,
with $T\gtrsim T_{0}$, where $T_{0}$ is defined by the infrared
energy cut-off of 10meV due to trigonal warping effects \cite{Zhang}.
In the same way, the renormalized DOS is $\rho(q)=[6\pi\gamma(q)]{}^{-1}/q\sim q^{-(1+\alpha_{1}/N)}$,
which can be measured directly on surfaces with scanning tunneling
spectroscopy experiments \cite{Li,Yankowitz}. 

In 2D systems, the electronic compressibility can be characterized
with single electron transistor measurements \cite{Martin}. By dimensional
analysis, the free electronic compressibility scales with temperature
as $\kappa\sim\gamma^{-2/3}T^{-1/3}$ \cite{Mahan}. In the same spirit,
interactions renormalize the scaling of the inverse compressibility,
\begin{equation}
\kappa^{-1}\sim T^{[1+2\alpha_{1}/(3N)]/3}\approx T^{1/3+0.1/N},\label{eq:kappa}
\end{equation}
which strongly deviates from the non-interacting result. 

In summary, we derived the effect of electron-electron interactions
in the renormalization of a variety of different physical observables
in the metallic phase of ABC graphene trilayers. 

We thank F. Guinea, V. N. Kotov and K. Mullen for valuable discussions.
B. U. acknowledges University of Oklahoma and NSF grant DMR-1352604
for partial support. 

\appendix
%dummy comment inserted by tex2lyx to ensure that this paragraph is not empty

\end{document}